\documentstyle[prd,aps,floats]{revtex}

\begin{document}
\preprint{astro-ph/9901268}
\draft

%
%
\input epsf
\renewcommand{\topfraction}{0.99}
\twocolumn[\hsize\textwidth\columnwidth\hsize\csname 
@twocolumnfalse\endcsname

\title{Critical collapse and the primordial black hole initial mass function} 
\author{Anne M.~Green}
\address{Astronomy Centre, University of Sussex, Brighton BN1 9QJ,~~U.~K.\\
and\\
Astronomy Unit, School of Mathematical Sciences, Queen Mary and
Westfield College,\\ Mile End Road, London, E1 4NS,~~U.~~K. (present address)}
\author{Andrew R.~Liddle}
\address{Astronomy Centre, University of Sussex, Brighton BN1 9QJ,~~U.~K.\\
and\\
Astrophysics Group, The Blackett Laboratory, Imperial College,
London, SW7 2BZ,~~U.~K. (present address)}  
\date{\today} 
\maketitle
\begin{abstract}
It has normally been assumed that primordial black holes (PBHs) always
form with mass approximately equal to the mass contained within the
horizon at that time.  Recent work studying the application of
critical phenomena in gravitational collapse to PBH formation has
shown that in fact, at a fixed time, PBHs with a range of masses are
formed.  When calculating the PBH initial mass function it is usually
assumed that all PBHs form at the same horizon mass. It is not clear,
however, that it is consistent to consider the spread in the mass of
PBHs formed at a single horizon mass, whilst neglecting the range of
horizon masses at which PBHs can form. We use the excursion set
formalism to compute the PBH initial mass function, allowing for PBH
formation at a range of horizon masses, for two forms of the density
perturbation spectrum.  First we examine power-law spectra with $n>1$,
where PBHs form on small scales. We find that, in the limit where the
number of PBHs formed is small enough to satisfy the observational
constraints on their initial abundance, the mass function approaches
that found by Niemeyer and Jedamzik under the assumption that all PBHs
form at a single horizon mass.  Second, we consider a flat
perturbation spectrum with a spike at a scale corresponding to horizon
mass $\sim 0.5 M_{\odot}$, and compare the resulting PBH mass function
with that of the MACHOs (MAssive Compact Halo Objects) detected by
microlensing observations. The predicted mass spectrum appears
significantly wider than the steeply-falling spectrum found
observationally.
\end{abstract}

\pacs{PACS numbers: 98.80.-k \hspace*{7.6cm}  astro-ph/9901268}

\vskip2pc]

\section{Introduction}

There has been a lot of interest recently in numerical studies of 
gravitational collapse, carried out for a wide range of matter models,
which appear to exhibit critical phenomena:
self-similarity, universality and power-law scaling of the black hole
mass (for a review and extensive references see e.g.~\cite{chop:rev}).
Consider a smooth one-parameter family of initial data described by a
parameter $p$, such that for $p>p_{{\rm c}}$ a black hole is formed and for
$p<p_{{\rm c}}$ no black hole is formed. Numerical simulations
(originally carried out by Choptuik for the case of a massless scalar
field~\cite{chop} and by Evans and Coleman~\cite{ec:rad} for a perfect
fluid with equation of state $p=\rho /3$) show
that the black hole mass, $M_{{\rm BH}}$, scales as
\begin{equation}
\label{mbh1}
M_{{\rm BH}} \propto (p-p_{\rm{c}})^{\gamma} \,,
\end{equation}
for $p \simeq p_{{\rm c}}$, where $\gamma$ is a universal scaling
exponent which is independent, for a given matter model, of the choice
of $p$ and the initial shape of the density fluctuations.

It has been pointed out by Niemeyer and Jedamzik~\cite{nj:crit} that
this work has an astrophysical application to primordial black hole
(PBH) formation.  For critical phenomena to occur, the initial
distribution of fluctuations must be such that most of the collapsing
fluctuations have magnitude not too far above the critical magnitude
for collapse.  This situation arises in the formation of PBHs from
density perturbations in the early universe, since the distribution of
large PBH forming fluctuations falls off rapidly, roughly as
$P({\delta}) \propto \exp(-\delta^2)$ where $\delta= \delta \rho
/\rho$. Generally this is not the case during astrophysical black
hole formation, such as stellar collapse where mass scales such as the
Chandrasekhar mass are introduced due to degeneracy pressure and other
effects violating the scale-free behaviour.\footnote{ Such physical
effects may also occur on small scales in the case of PBH formation so
that arbitrarily small mass PBHs will not be formed, but this does not
affect the general validity of Eq.~(\ref{mbh1}).}

Niemeyer and Jedamzik \cite{n:crit} have investigated the evolution of
spherically-symmetric perturbations of the energy density in
unperturbed Hubble flow during radiation domination\footnote{Realistic
PBH formation differs from the study of perfect fluid collapse in
Ref.~\cite{ec:rad}, where the initial data was embedded in an
asymptotically flat space-time, in that the background spacetime is
Friedmann--Robertson--Walker.}  for three different perturbation
shapes at horizon crossing: a gaussian overdensity, which tends to the
background Friedmann--Robertson--Walker metric at infinity, a Mexican
Hat function, and an unspecified fourth-order polynomial.  In the
latter two cases the inner overdensity is compensated by a surrounding
underdense region.  The initial conditions are tuned to
super-criticality by adjusting the amplitude of the perturbations.
Their simulations confirm the existence of the mass scaling relation
in the case of PBH formation:
\begin{equation}
\label{mbh}
M_{{\rm BH}} = k M_{\rm{H}}\left( \delta - \delta_{\rm{c}} \right)
        ^{\gamma} \,,
\end{equation}
where $M_{\rm{H}}$ is the horizon mass at the time the fluctuation
enters  the horizon, $\delta$ is defined as the additional mass
inside the horizon radius in units of the horizon mass and $\gamma
\approx 0.37$ for all three shapes of perturbation. For gaussian shaped
fluctuations $k=11.9$ and $\delta_{\rm{c}}=0.70$, for the 
Mexican Hat fluctuations $k=2.85$ and $\delta_{\rm{c}}=0.67$, and for the
fourth-order polynomial $k=2.39$ and $\delta_{\rm{c}}=0.71$, suggesting
that $\delta_{\rm{c}} \sim 0.7$ for all perturbation shapes. This value
of $\delta_{{\rm c}}$ is about a factor of $2$ larger than the value
found analytically~\cite{pbhf:carr} by requiring that the over-dense
region exceeds its Jeans length at the time that it stops expanding.

The formation, at a fixed horizon mass, of PBHs with a range of masses
is a significant change from the standard picture where it is assumed
that all PBHs are formed with $M_{\rm{BH}} \approx
M_{\rm{H}}$~\cite{pbhf:ch,pbhf:carr}. Niemeyer and Jedamzik have
determined the PBH initial mass function under the assumption that all
PBHs form at the same horizon mass. This assumption is often used; for
example, with a power-law spectrum with spectral index greater than
one it is assumed that the vast majority of the black hole formation
occurs at the shortest possible scale. The usual justification is that
PBH formation is extremely sensitive to the amplitude of the
perturbations, and hence inherits a strong dependence on scale even if
the variation of the spectrum is weak. It is not clear, however, that
it is consistent to consider the spread in the mass of PBHs formed at
a single horizon mass, whilst neglecting the range of horizon masses
at which the PBHs are formed. Investigation of this issue is the main
purpose of this paper. In the end, we shall in fact find that this
assumption works very well in the cases of astrophysical interest.

Heuristically, to find the mass of a PBH formed at a given point in
space we need to smooth the density field at that point, $\delta({\bf
x})$, on a range of mass scales, to produce $\delta(M)$, where $M$ is
the smoothing scale, and then evaluate each $\delta(M)$ at the time
when that scale crosses the horizon, $M=M_{{\rm H}}$.  If
$\delta(M=M_{{\rm H}})>\delta_{{\rm c}}$ then a PBH is formed, with
mass given by Eq.~(\ref{mbh}), at that horizon mass.  However, that
condition will be satisfied for a range of smoothing masses; the
actual mass of the PBH formed at the point ${\bf x}$ is given by the
largest value of $M_{{\rm BH}}$ found as the smoothing scale is
decreased, which will not be the largest smoothing scale
giving a density contrast above the threshold due to the dependence of
the mass on $(\delta-\delta_{{\rm c}})$.  To find the optimal
smoothing mass, we use the excursion set formalism \cite{bceketc}
which was introduced in large-scale structure studies to determine the
mass function, merger rates and clustering of collapsed objects
e.g.\cite{bceketc,pmlc:clus}.

\section{The excursion set formalism}
\subsection{Standard formalism}

To examine the density perturbations on a given scale we must smooth 
the density field, as described above, using a window function 
$W(R_{{\rm f}}, |{\bf x}
- {\bf x}'|)$ with radius $R_{{\rm f}}$.  The density contrast is defined as 
$\delta(\bf{x}) = (\rho(\bf{x})-\bar{\rho})/\bar{\rho}$, 
and the smoothed 
version is given by:
\begin{eqnarray}
\label{dels}
\delta(R_{{\rm f}}, {\bf x})& = & \int^{\infty}_{0} W(R_{{\rm f}} ,
        |{\bf x}-{\bf x}'|) \delta({\bf x}') {\rm d}^3 {\bf x}' \nonumber \\
         & =& \frac{1}{(2 \pi)^3} \int^{\infty}_{0} W(kR_{{\rm f}}) 
        \delta({\bf k}) \exp{(-i {\bf k}.{\bf x}})
        {\rm d}^3 {\bf k} \,,
\end{eqnarray}  
where $W(k R_{{\rm f}})$ and $\delta({\bf k})$ are the
Fourier transform of the window function and the unsmoothed density
contrast respectively, with $k = |{\bf k}|$. The variance of the mass 
distribution
$\sigma^2(R_{{\rm f}})$, as defined in \cite{LL}, is given by
\begin{equation}
\label{sig}
\sigma^{2}(R_{{\rm f}}) =\int^{\infty}_0 W(kR_{{\rm f}})
  {\cal P}_{\delta}(k) \frac{{\rm d} k}{k} \,,
\end{equation}
where ${\cal P}_{\delta}(k)$ is the spectrum of $\delta$ 
\begin{equation}
{\cal P}_{\delta}(k)= \frac{k^3}{2 \pi^2} \, \langle |
        \delta({\bf k})|^2 \rangle \,.
\end{equation}

The effect of varying $R_{{\rm f}}$, {\it at fixed time}, can be found by 
differentiating
Eq.~(\ref{dels})~\cite{pmlc:clus}:
\begin{eqnarray}
\label{lang}
\frac{ \partial \delta(R_{{\rm f}},{\bf x})}{\partial R_{{\rm f}}}& = & 
      \frac{1}{(2 \pi)^3} \int^{\infty}_{0} \delta({\bf k})
      \frac{ \partial W(kR_{{\rm f}})}{\partial R_{{\rm f}}}
      \\ \nonumber && \times
     \exp{(-i{\bf k}.{\bf x})} {\rm d}{\bf k} \equiv \eta(
      R_{{\rm f}},{\bf x}) \,.
\end{eqnarray}
This has the form of a Langevin equation; the change in $\delta
(R_{{\rm f}},{\bf x})$ when $R_{{\rm f}}$ is changed is given in terms
of a stochastic force $\eta(R_{{\rm f}},{\bf x})$, which depends on the form
of the window function used.  

A particularly good choice of window function if one wants to make 
analytical progress is the
sharp $k$-space window function  
\begin{equation}
\tilde{W}(kR_{{\rm f}})= \Theta(1-kR_{{\rm f}}) \,.
\end{equation}
Its strength is that the only effect of decreasing the smoothing
radius is to add new Fourier modes of the unsmoothed density contrast
$\delta({\bf x})$ to the integral for $\sigma^2$. For gaussian
perturbations, as we will be assuming throughout,\footnote{The
assumption of gaussianity has been challenged for large PBH-forming
fluctuations~\cite{gauss:var}; however it is reasonable to maintain
this assumption in order to assess the effect of critical collapse on
the PBH mass function.} these new modes are uncorrelated with the ones
already included in the integral on a larger smoothing scale. The
change in $\delta(R_{{\rm f}},{\bf x})$ caused by the new modes is
therefore independent of its value on the larger smoothing scale,
leading to a random walk without memory. For other choices of window
function, including the commonly-used top-hat and gaussian forms, this
nice property doesn't hold since changing $R_f$ alters the
contribution to $\sigma^2$ from modes of {\em all} wavenumbers. That
said, although the sharp $k$-space window function is calculationally
advantageous, one wouldn't expect physical results to be all that
dependent on the choice of smoothing.

With the sharp $k$-space window function, Eq.~(\ref{lang}) 
simplifies to 
\begin{equation}
\label{rw1}
\frac{\partial \delta({\bf x}, \Lambda)}{\partial \Lambda} = \zeta(
         \Lambda) \,,
\end{equation}
independent of position, where $\Lambda \equiv \sigma^2(R_{{\rm f}})$ 
is a pseudo-time variable and 
\begin{equation}
\langle \zeta( \Lambda_{1}) \zeta(\Lambda_{2}) \rangle=
\delta_{{\rm D}}(\Lambda_{1}-\Lambda_{2}) \,,
\end{equation}
the right-hand side being the Dirac delta function. 
 
Eq.~(\ref{rw1}) can be integrated to give
\begin{equation}
\label{rw2}
\delta(\Lambda+\gamma)- \delta(\Lambda) = \sqrt{\gamma}\,  G
\end{equation}
where $G$ is a gaussian random variable with mean zero and unit
variance, and since this equation is exact the step size, $\gamma$,
can be chosen freely.  Stochastic processes which are governed by this
equation, such as self diffusion in a hard sphere gas, are known as
Wiener processes. We can think of the values of $\delta(\Lambda)$
produced as $\Lambda$ is increased (or equivalently $R_{{\rm f}}$
decreased) as mapping out a trajectory, analogous to the path of a
self-diffusing particle.  Each trajectory has the initial condition
$\delta(0)=0$ since in the limit that $R_{{\rm f}} \rightarrow
\infty$, $\delta(R_{{\rm f}}, {\bf x}) \rightarrow 0$ by definition of
the mean density.

Using Eq.~(\ref{rw2})
and a random number generator, we can generate an ensemble of 
trajectories each representing the variation, with smoothing scale,
of the smoothed density field at a different point in space.
The trajectories of $\delta(\Lambda)$
produced for a chosen range of $\Lambda$ values are independent of the form of 
the power spectrum. To relate
each value of $\Lambda$ to a mass scale, $M$, we need to choose a form
for the power spectrum and then use Eq.~(\ref{sig}) to find the 
relationship between $\Lambda$ and $M$.

\subsection{Application to PBH formation}
        
The trajectories give the variation of $\delta(M)$ at fixed time;
however, the condition for PBH formation on a given scale,
$\delta(M)>\delta_{{\rm c}}$, and Eq.~(\ref{mbh}) which gives the mass
of the PBH produced, apply when that scale crosses the horizon, i.e.
$M=M_{{\rm H}}$. We must therefore choose a fixed time at which to
generate our trajectories, and then evolve each value of $\delta(M)$
forwards, or backwards, in time to the epoch at which that scale
crosses the horizon. During radiation domination perturbations which
are outside the horizon grow as~\cite{LL}
\begin{equation}
\delta \propto t \propto M_{{\rm H}} 
\end{equation}
and therefore 
\begin{equation}
\delta_{{\rm hc}}(M_{{\rm H}}) = \delta_{{\rm ft}}(M_{{\rm H}}) 
           \left( \frac{M_{{\rm H}}}
              {M_{{\rm H, min}}} \right) \,,
\end{equation}
where `hc' and `ft' denote quantities evaluated at horizon crossing and
the chosen fixed time respectively. For PBH formation we require 
$\delta_{{\rm ft}}(M_{{\rm H}}) > \delta_{{\rm c, ft}}(M_{{\rm H}})$ where
\begin{equation}
\delta_{{\rm c, ft}}(M_{{\rm H}}) = \delta_{{\rm c}} \left( 
             \frac{M_{{\rm H}}}{M_{{\rm H, min}}}
           \right) \,.
\end{equation}

\section{Power-law power spectra}

We first examine the case where the primordial power spectrum is a power-law
\begin{equation}
{\cal P}(\delta) \propto k^{n+3} \,,
\end{equation}
[equivalently $P(k) \propto k^{n}$, where $P(k)= \langle | 
\delta_{\bf k}|^2 \rangle$ and $n$ is known as the spectral index]. 
Inserting this form for the power spectrum into 
Eq.~(\ref{sig}) gives $\sigma^2(R) \propto R^{-(n+3)}$. 

During radiation domination the mass in a comoving region evolves,
reducing as $1/a$. Ultimately we are interested in the mass associated
with a given scale when that scale crosses the horizon; horizon
crossing is given by $R \propto 1/aH \propto t^{1/2} \propto M_{{\rm
H}}^{1/2}$ \cite{gl} so that, still at fixed time,
\begin{equation}
\sigma^2(M_{{\rm H}}) = \sigma^2(M_{{\rm H, min}}) \left( \frac{M_{{\rm H}}}
           {M_{{\rm H, min}}} \right)^{-(n+3)/2} \,.
\end{equation}
where we have chosen our fixed time as the epoch immediately after the 
reheating period at the end of inflation when the horizon mass has its
minimum value, $M_{{\rm H, min}}$, which is determined by the reheat 
temperature, $T_{{\rm RH}}$, via
\begin{equation}
M_{{\rm H,min}} = M_{{\rm H,0}} \left( \frac{T_{{\rm eq}}}{T_{{\rm
                 RH}}}\right) ^{2} \left( \frac{T_{{\rm 0}}}
                 {T_{{\rm eq}}} \right)^{3/2} \,.
\end{equation}
We can obtain $\sigma(M_{{\rm H,min}})$ from the mass variance on the 
present horizon scale,
using the variation with $M$ of the mass variance at horizon crossing
(for details see Ref.~\cite{gl}):
\begin{equation}
\sigma(M_{{\rm H, min}}) = \sigma(M_{{\rm H, 0}}) \left( \frac{M_{{\rm 
          eq}}}{M_{0}} \right)^{(1-n)/6} \left( \frac{ M_{{\rm H,min}}}
         {M_{{\rm eq}}} \right)^{(1-n)/4} \,,
\end{equation}
where `0' and `eq' denote quantities
evaluated at the present day and matter--radiation equality, and 
$\sigma(M_{{\rm H, 0}}) =9.5 
\times 10^{-5}$ using the COBE normalization~\cite{gl}.

There are a number of well-known constraints on the abundance of PBHs
over a wide range of mass scales; for an up-to-date review see
Ref.~\cite{pbh:rev}.  Typically only of order $10^{-20}$ of the energy
density of the universe, at the time that they form, can go into PBHs.
This leads to the constraint that $n$ must be less than about 1.25
\cite{gl}, with some dependence on the value of $M_{{\rm H,min}}$.  It
is obviously not feasible to run simulations where only one trajectory
in $10^{20}$ produces a PBH.  Using larger values of $n$, around
$1.3$, a manageable number of trajectories will form PBHs but the
values of $\delta_{{\rm hc}}(M_{{\rm H}})$, and hence $M_{{\rm BH}}$
produced will typically be larger than when $n \sim 1.25$.  We can
however examine how the distribution of PBH masses behaves as $n$ is
decreased and compare its limiting behaviour with the mass function
found analytically by Niemeyer and Jedamzik under the assumption that
all PBHs form at the same horizon mass\cite{nj:crit}
\begin{eqnarray}
\label{njmf}
\frac{{\rm d} N}{{\rm d} ( \ln{M_{{\rm BH}}})}& =& \frac{1}
         {\sqrt{2 \pi}  \sigma \gamma} \left( \frac{M_{{\rm BH}}}
         {k M_{{\rm H}}} \right)^{1/\gamma} \nonumber \\
       &&  \times  \exp{ \left( - \frac{
         \left( \delta_{{\rm c}}+(M_{{\rm BH}}/ k M_{{\rm H}})^{1/\gamma} 
         \right)^2}{2 \sigma^2} \right)} \,,
\end{eqnarray}
where ${\rm d} N$ is the number of PBHs per logarithmic mass interval
${\rm d}(\ln{M_{{\rm BH}}})$, and $\sigma \equiv \sigma(M_{{\rm H}})$.
The Niemeyer and Jedamzik mass function (NJMF) will be valid if the
distribution of the horizon masses at which the PBHs are formed is
close to a delta function.

\begin{figure}[t]
\centering
\leavevmode\epsfysize=6.3cm \epsfbox{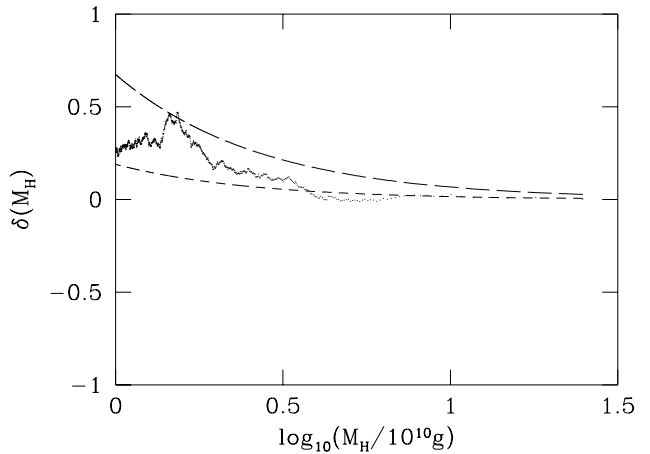}\\
\caption[traj1]{\label{traj1} A typical PBH forming trajectory resulting
from a power-law spectrum with $n=1.3$. The dotted points are the
values of $\delta(M_{{\rm H}})$, the long-dashed line shows the threshold, 
at the
fixed time for PBH formation $\delta_{{\rm c, ft}}(M_{{\rm H}})$, and the 
short-dashed 
line shows the typical size of perturbations on each scale, 
$\sigma(M_{{\rm H}})$.}
\end{figure}

As an example we choose $M_{{\rm H,min}}= 10^{10}$g, which corresponds
to $T_{{\rm RH}} \approx 3 \times 10^{11}$ GeV, and in Eq.~(\ref{mbh})
we use the values of $k$ and $\delta_{{\rm c}}$ from
Ref.~\cite{n:crit} for Mexican hat shaped fluctuations. Our general
conclusions will however be independent of these choices.  Since the
probability of PBH formation falls off rapidly with increasing horizon
mass we generate trajectories using 1000 equally spaced steps in
$\Lambda$, giving adequate coverage of the range of horizon masses at
which PBHs form. Fig.~\ref{traj1} shows a typical PBH forming
trajectory with $n=1.3$.

\begin{figure}[t]
\centering
\leavevmode\epsfysize=6.3cm \epsfbox{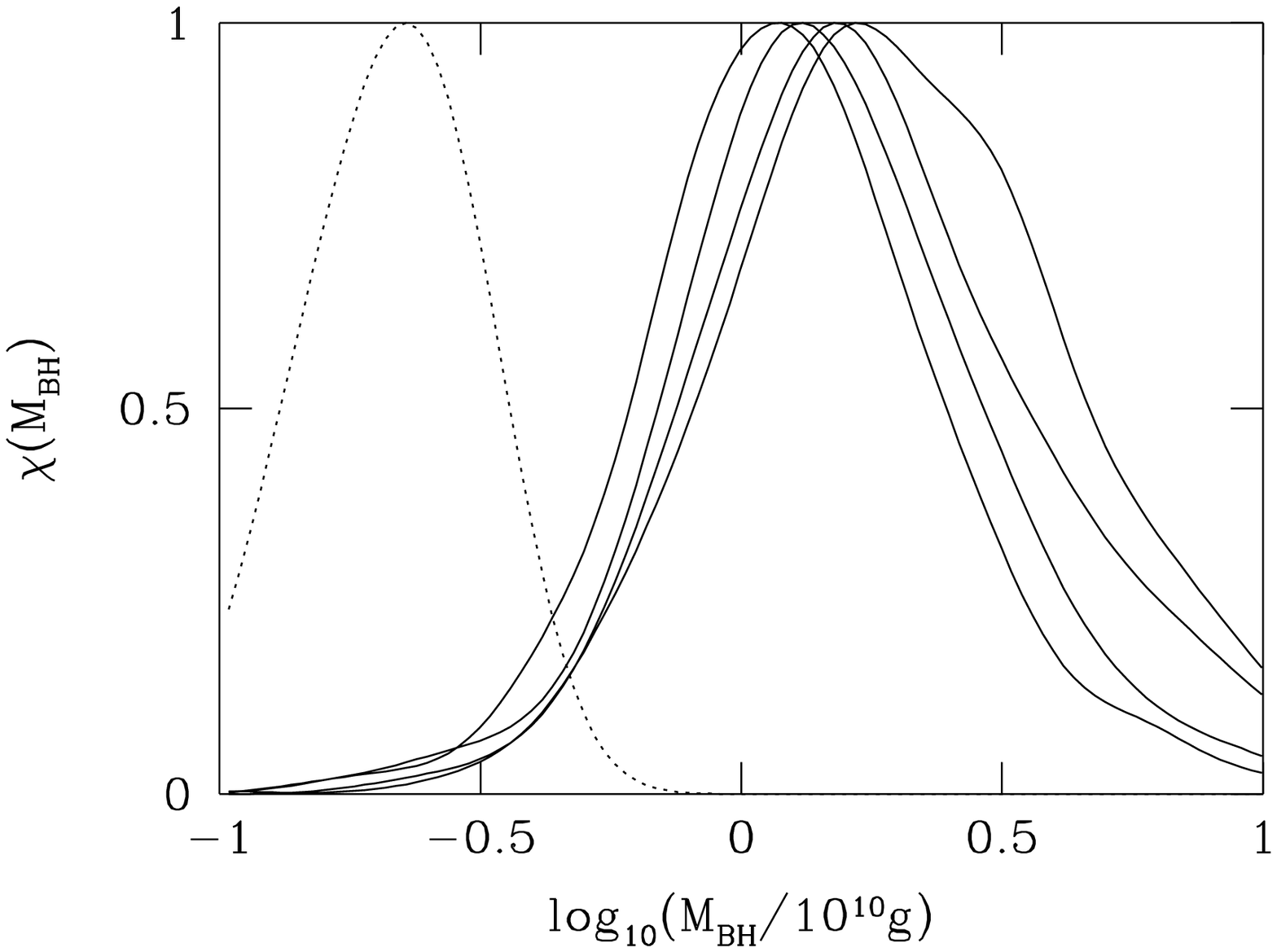}\\
\leavevmode\epsfysize=6.3cm \epsfbox{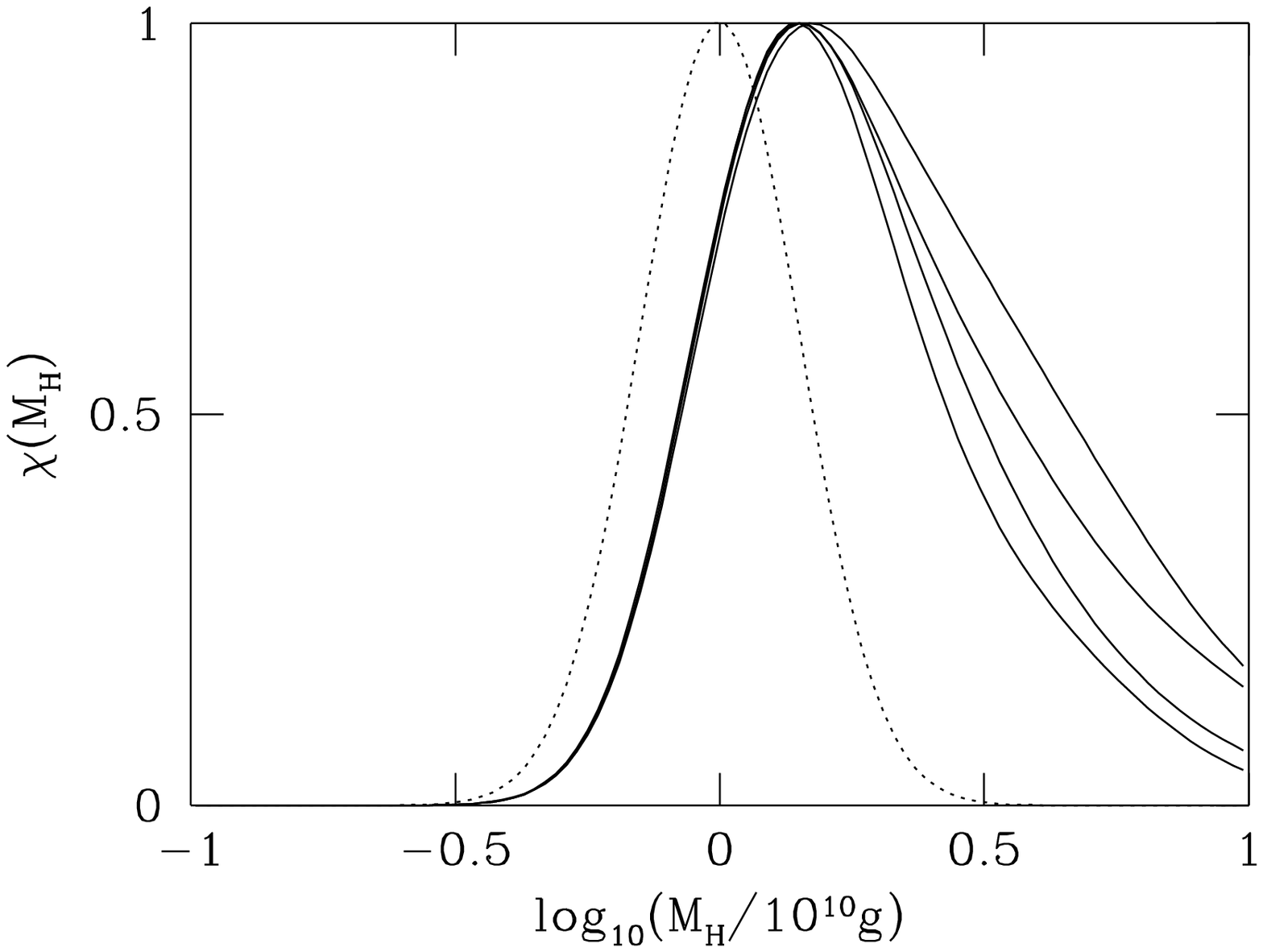} \\
\caption[mbhfig]{\label{mbhfig} The smoothed distributions of the
actual PBH masses, $\chi(M_{{\rm BH}})$, (top panel) and the horizon
masses at their formation, $\chi(M_{{\rm H}})$, (lower panel). From
top to bottom on the right-hand side of each diagram, the solid lines
are for $n=1.310$, 1.305, 1.300 and 1.295. In the top panel the dotted
line shows the Niemeyer and Jedamzik mass function, evaluated for
$\sigma(10^{10}{\rm g})=0.032$ and smoothed on the same scale, while
in the bottom panel the dotted line is a smoothed delta function,
centered on $M_{{\rm H}}=10^{10}$g.}
\end{figure}

We ran simulations producing 1000 PBHs for each of $n$=1.310, 1.305,
1.300, and 1.295 and in each case found the distributions (smoothed
with a gaussian to remove the effects of discreteness from the finite
number of PBHs) of the PBHs masses and the horizon masses at which
they formed. We denote these as $\chi(M_{{\rm BH}})$ and $\chi(M_{{\rm
H}})$ respectively, so that the number of PBHs per logarithmic mass
interval produced in our simulations is given by $\chi(M_{{\rm
BH}}){\rm d}(\ln{M})$.  The upper panel of Fig.~\ref{mbhfig} shows the
smoothed mass distributions $\chi(M_{{\rm BH}})$, along with the
Niemeyer and Jedamzik mass function (smoothed on the same scale to
allow direct comparison) evaluated for $M_{{\rm H}}=10^{10}$g and
$\sigma=0.032$, the value of $\sigma(10^{10}{\rm g})$ for $n=1.23$
(the maximum value of $n$ allowed for $M_{{\rm H,min}}= 10^{10}$ g by
the observational constraints). We normalize all mass distributions to
$1$ at their peak to facilitate comparison.  As $n$ is decreased, the
PBH mass function tends towards that found by assuming that all PBHs
form at a single horizon mass. We can test this further by
examining the {\em horizon} masses at which the PBHs form, rather than
their actual masses.  The lower panel shows the smoothed distributions
of the horizon masses at which the PBHs formed, plotted along with a
delta function centered on $M_{{\rm H}}=10^{10}$g and smoothed on the
same scale. As $n$ is decreased the horizon mass distribution narrows. 

We therefore conclude that for power-law spectra, with slope
satisfying the constraints from the observational limits on PBH
abundance, the assumption that all PBHs form at the same horizon mass
is a good approximation. Yokoyama~\cite{yok} has re-evaluated the
constraints on the initial mass fraction of PBHs, taking account of
the mass scaling relation under this assumption.  He found that the
constraints on the initial abundance of PBHs are unchanged if
formation occurs at $M_{{\rm H}}< 5 \times 10^{14}$g.  The limit on
the abundance of PBHs with $M_{{\rm BH}}= 5 \times 10^{14}$g, which
are evaporating today, coming from the $\gamma$-ray background, now
leads to stronger limits on the initial abundance if $5 \times 10^{14}
{\rm g} < M_{{\rm H}} < 10^{17}$g, as for horizon masses in this range
the low-mass tail of the distribution has a significant population
light enough to be evaporating today.

\section{Spiky power spectra}

Observations of microlensing of stars in the Large Magellanic
Cloud~\cite{ma:cho} appear to show that a large fraction of the halo
of the galaxy is in the form of MAssive Compact Halo Objects (MACHOs)
with masses of $M \sim 0.5 M_{\odot}$.  It has been proposed that
MACHOs may be PBHs formed either during the QCD
epoch~\cite{qcd:pbh1,qcd:pbh2} because of the reduction in pressure
forces at that time, or due to a spike in the primordial density
perturbation spectrum~\cite{ss:pbh} at the scale corresponding to
$M_{{\rm H}} \sim 10^{33}$g.

The mass distribution of the MACHOs is steeply peaked. In
Ref.\cite{ma:cho} the observed microlensing events are fitted with
power-law mass functions:
\begin{eqnarray}
\label{machdist1}
\psi(M)&=& A M^{\alpha} \,\,\, (M_{{\rm min}} < M < M_{{\rm max}}) \\ 
      &=& 0 \,\,\,  ({\rm otherwise}) \nonumber \,,
\end{eqnarray}
where $\psi(M){\rm d}M$ is the mass fraction of MACHOs between $M$ and
$M + {\rm d}M$ and A is determined by the mass fraction of the halo in
the form of MACHOs. The maximum likelihood fit (for their 8 event
sample) is found for $\alpha=-3.9$ and $M_{{\rm min}}=0.30
M_{\odot}$.\footnote{ For a subset of 6 events (chosen to exclude a
binary lensing event whilst maintaining the mean duration) the most
likely mass distribution is a delta function.}

The spread in the masses of PBHs formed at a single horizon mass,
found when critical collapse is taken into account, raises the
question as to whether or not it is possible to produce a population
of PBHs with a sufficiently sharply peaked mass distribution. There
are several models of inflation which produce a spike in the power
spectrum on a particular scale~\cite{ss:pbh}.  For generality, we will
take the spectrum of the curvature perturbations, ${\cal P}_{{\cal
R}}$, which is defined, analogously to ${\cal P}_{\delta}$,
as~\cite{LL}
\begin{equation}
{\cal P}_{{\cal R}} =  \frac{k^3}{2 \pi^2} \, 
            \langle | {\cal R}({\bf k}) | \rangle \,,
\end{equation}
to have the form of a gaussian spike at a scale $k_{{\rm c}}= k_{{\rm
eq}}(M_{{\rm H, eq}}/M_{{\rm H, c}})^{0.5}= 4.25 \times 10^{10} \, {\rm
Mpc}^{-1}$, corresponding to $M_{{\rm H, c}} = 10^{33}$g, with
variable amplitude and width, $\Sigma$, superimposed on a flat
spectrum normalized to the COBE data on the present horizon scale. The
COBE normalization gives ${\cal P}_{{\cal R}}=2.28 \times 10^{-9}$
\cite{LL} so our spectrum is
\begin{equation}
{\cal P}_{{\cal R}} = 2.28 \times 10^{-9} + \frac{C}{\sqrt{2 \pi} \Sigma}
         \exp{ \left(- \frac{\left(k-k_{{\rm c}}\right)^2}
          { 2 \Sigma^2} \right)} \,,
\end{equation}

\begin{figure}[t]
\centering
\leavevmode\epsfysize=6.3cm \epsfbox{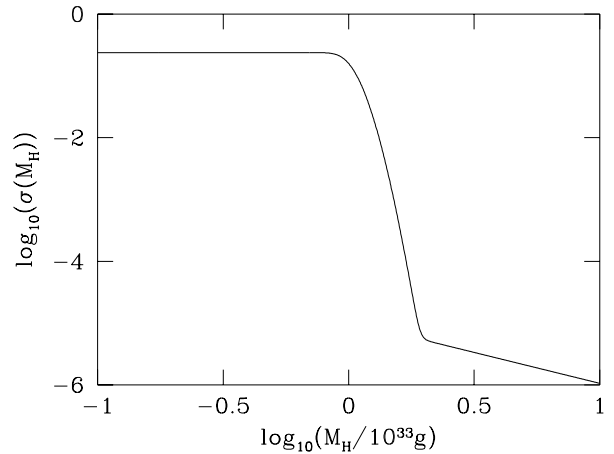} \\
\caption[sigm]{\label{sigm} The form of $\sigma(M_{{\rm H}})$ produced 
by a spike in the density perturbation spectrum with $C= 1 \times 10^{10}
{\rm Mpc}^{-1}$ and $\Sigma= 2 \times 10^{9}{\rm Mpc}^{-1}$.}
\end{figure}

In practice the results from taking a spectrum of this form are quite
general, as the bulk of the PBH production occurs right at the peak of
the power spectrum. A Taylor expansion in the vicinity of the peak
requires only the amplitude and curvature, and the two parameters of
the gaussian can reproduce an arbitrary function near the peak.  Our
results will therefore prove extremely general.

With this more complicated spectrum, in order to obtain $\Lambda(M)
\equiv \sigma^2(M)$ we have to numerically integrate Eq.~(\ref{sig})
for each ${\cal P}_{{\cal R}}$ we choose, where during radiation
domination
\begin{equation}
\delta({\bf k}) = \frac{4}{9} \left( \frac{k}{aH} \right)^2 
            \cal{R}\sl({\bf k}) \,,
\end{equation}
so that
\begin{equation}
{\cal P}_{\delta}= \frac{16}{81} \left( \frac{k}{aH} \right)^4 
      {\cal P}_{{\cal R}} \,,
\end{equation}
Fig.~\ref{sigm} shows $\sigma(M_{{\rm H}})$ when 
$C= 1 \times 10^{10}\,{\rm Mpc}^{-1}$ and $\Sigma= 2 
\times 10^{9}\,{\rm Mpc}^{-1}$.

\begin{figure}[t]
\centering
\leavevmode\epsfysize=6.3cm \epsfbox{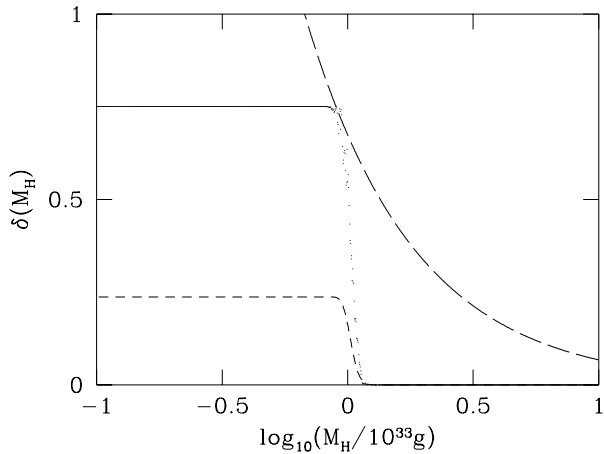}\\
\caption[trajss]{\label{trajss} A PBH forming trajectory produced when
$\cal{P}_{\cal{R}}\sl$ has a gaussian spike with $\Sigma=10^{9} {\rm
Mpc}^{-1}$ and $C=10^{10} {\rm Mpc}^{-1}$. The amplitude jumps
dramatically when the spike is encountered. The dotted points are the
values of $\delta(M_{{\rm H}})$, the long-dashed line shows the
threshold $\delta_{{\rm c, ft}}$ for PBH formation,
and the short-dashed line shows the typical size of perturbations on
each scale $\sigma(M_{{\rm H}})$.}
\end{figure}

We have run simulations for a range of values of $C$ and $\Sigma$
producing, in each case, 1000 PBHs, choosing the fixed time at which
the trajectories are produced as the epoch when $M_{{\rm
H}}=10^{33}$g, corresponding to half a solar mass.  Density
perturbation spectra with the same value of $C$ but different $\Sigma$
have the same value of $\sigma(M_{{\rm H}})$ on small scales ($M_{{\rm
H}}<10^{33}$g), decreasing as $C$ is increased.  Once again it is not
feasible to use spectra which produce PBH abundances consistent with
the observational limits.  We therefore choose values of $C$ and
$\Sigma$ which lead to larger values of $\sigma(M_{{\rm H}}) $ on
small scales ($\sim 0.2$) and examine the behaviour of the PBH mass
and horizon mass distributions at formation as $C$ and $\Sigma$ are
varied.  Fig.~\ref{trajss} shows an example of a PBH forming
trajectory with $\Sigma=10^{9}{\rm Mpc}^{-1} $ and $C= 10^{10} \, {\rm
Mpc}^{-1}$.

\begin{figure}[t]
\centering
\leavevmode\epsfysize=6.3cm \epsfbox{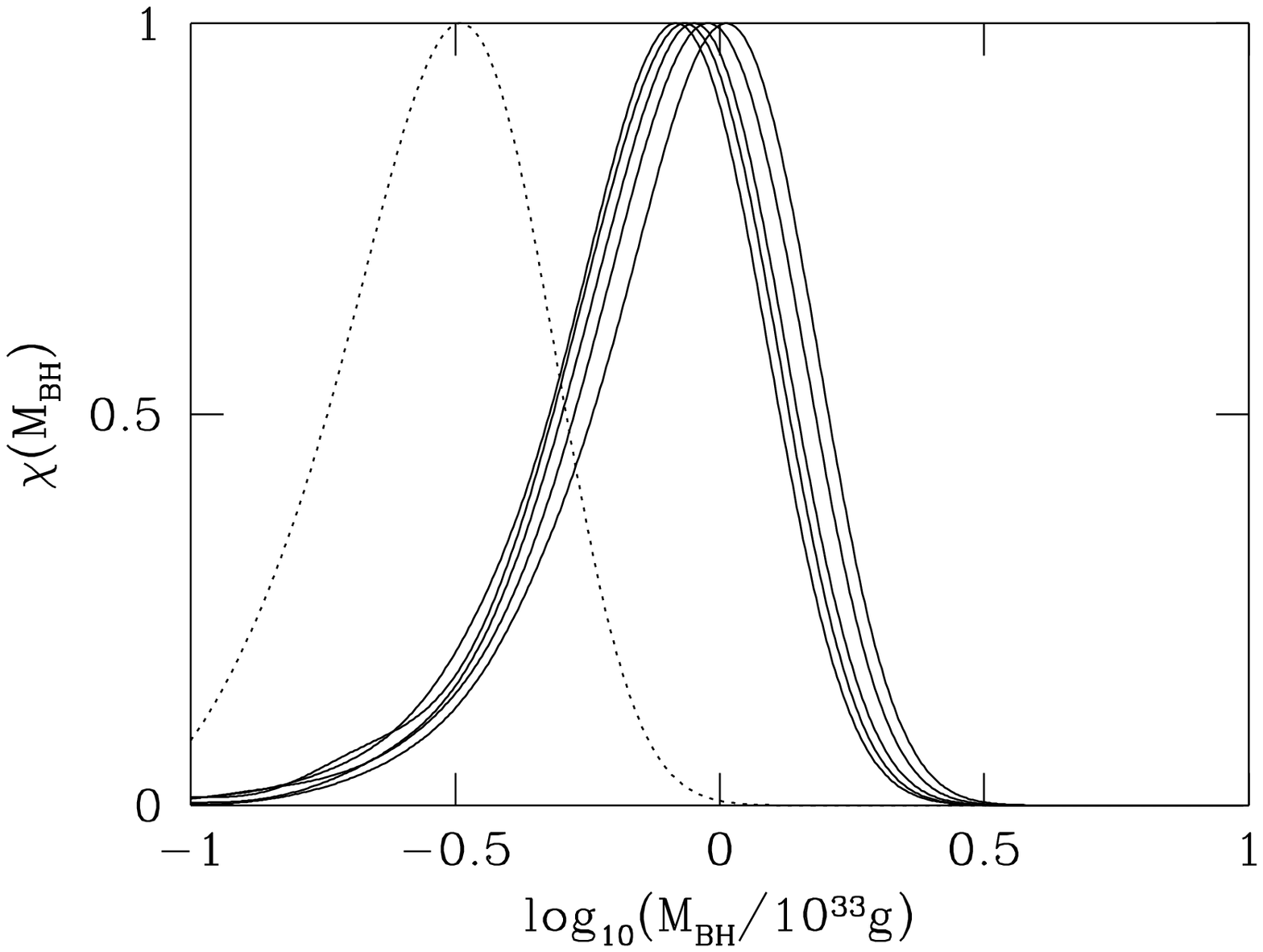}\\
\leavevmode\epsfysize=6.3cm \epsfbox{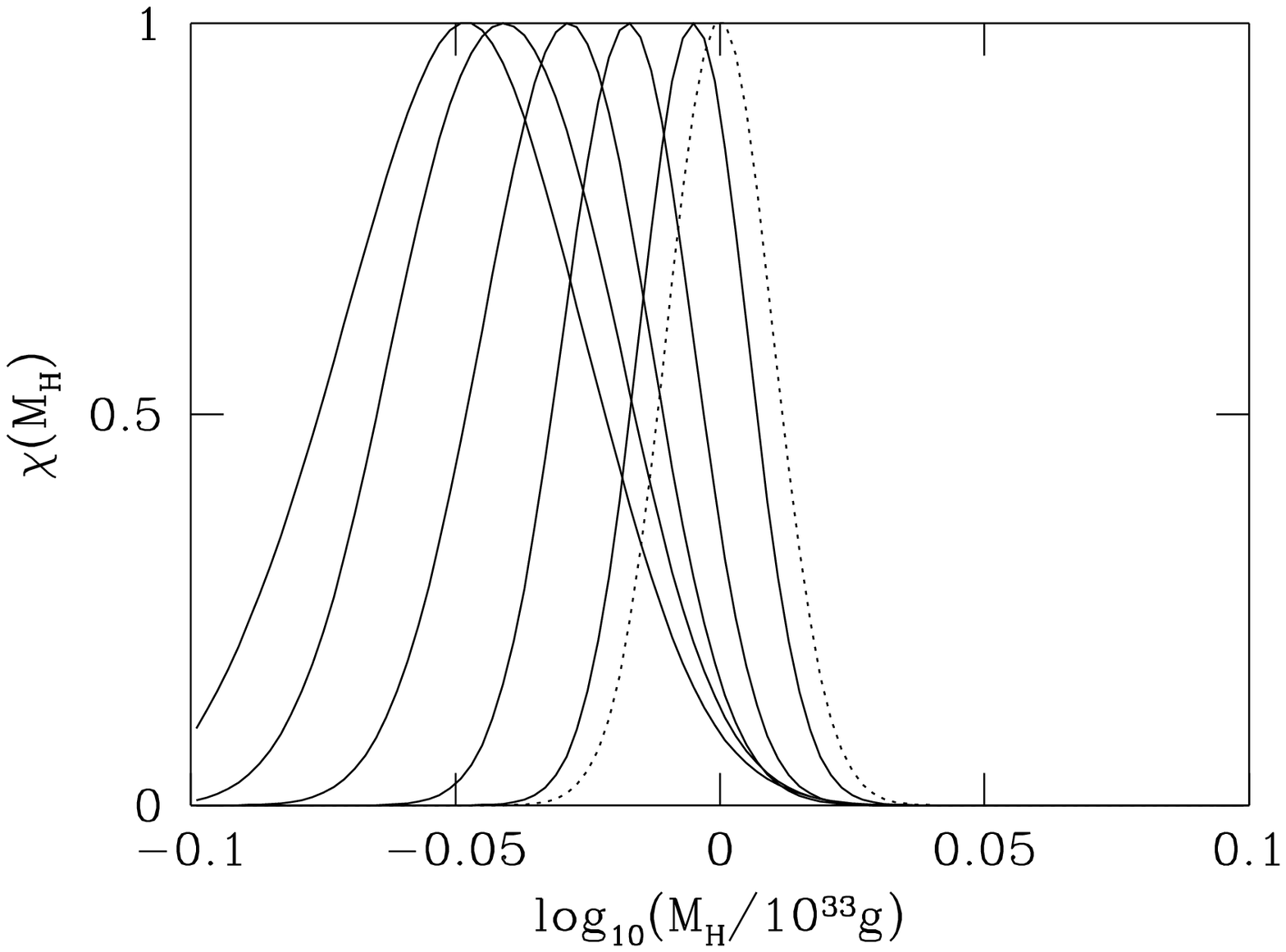}\\
\caption[Cfixmbh]{\label{Cfixmbh} The smoothed distribution of the
actual masses, $\chi(M_{{\rm BH}})$, (upper panel) and horizon masses,
$\chi(M_{{\rm H}})$, (lower panel) of PBHs formed due to a spike in
the density perturbation spectrum with $C= 1 \times 10^{10} \,{\rm
Mpc}^{-1}$.  From right to left on the right-hand side of each
diagram, the solid lines show $\Sigma= 1 \times 10^{8}, 5 \times
10^{8}, 1 \times 10^{9}, 1.5 \times 10^{9}$ and $2 \times10^{9} {\rm
Mpc}^{-1}$.  The dotted line in the upper panel shows the smoothed
Niemeyer and Jedamzik mass function evaluated for $\sigma(10^{33}{\rm
g})=0.053$, and in the lower panel a smoothed delta function centered
on $M_{{\rm H}}=10^{33}$g.}
\end{figure}

The upper panel of Fig.~\ref{Cfixmbh} shows the smoothed distribution
of the PBH masses, $\chi(M_{{\rm BH}})$, for $C=1 \times 10^{10} \, {\rm 
Mpc}^{-1}$ and
a range of values of $\Sigma$.  For comparison we also plot the
smoothed NJMF evaluated for $\sigma(M_{{\rm H}}=10^{33}{\rm
g})=0.053$, the largest value compatible with the requirement that the
present-day density of PBHs is consistent with the present-day age and
expansion rate of the universe, i.e.  $\Omega_{{\rm PBH,0}}< 1$.  For
all the values of $\Sigma$ we used, the PBH mass function has the same
shape as the NJMF, but it is centered on larger values of $M_{{\rm
BH}}$, shifting towards smaller values of $M_{{\rm BH}}$ as $\Sigma$
is increased.  The lower panel shows the smoothed distribution of the
horizon masses at which the PBHs form, $\chi(M_{{\rm H}})$, for the
same set of $C$ and $\Sigma$ values, along with a smoothed delta
function centered on $M_{{\rm H}}=1 \times 10^{33}$g.  The horizon
mass distributions have the same shape as the delta function but are
centered at smaller $M_{{\rm H}}$.  As $\Sigma$ is decreased the
centre of the distribution tends towards $M_{{\rm H}}=10^{33}$g.

\begin{figure}[t]
\centering
\leavevmode\epsfysize=6.3cm \epsfbox{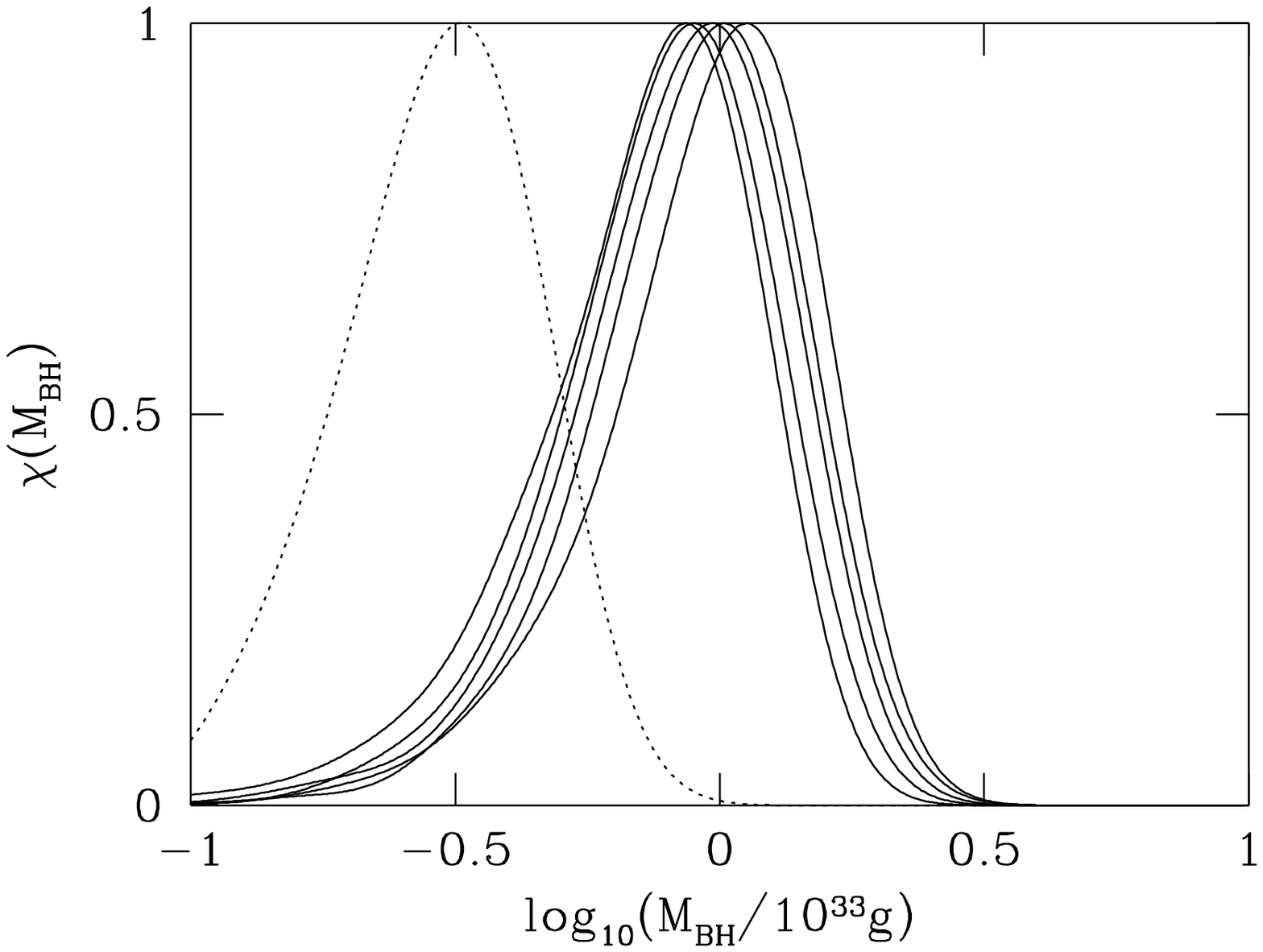}\\
\leavevmode\epsfysize=6.3cm \epsfbox{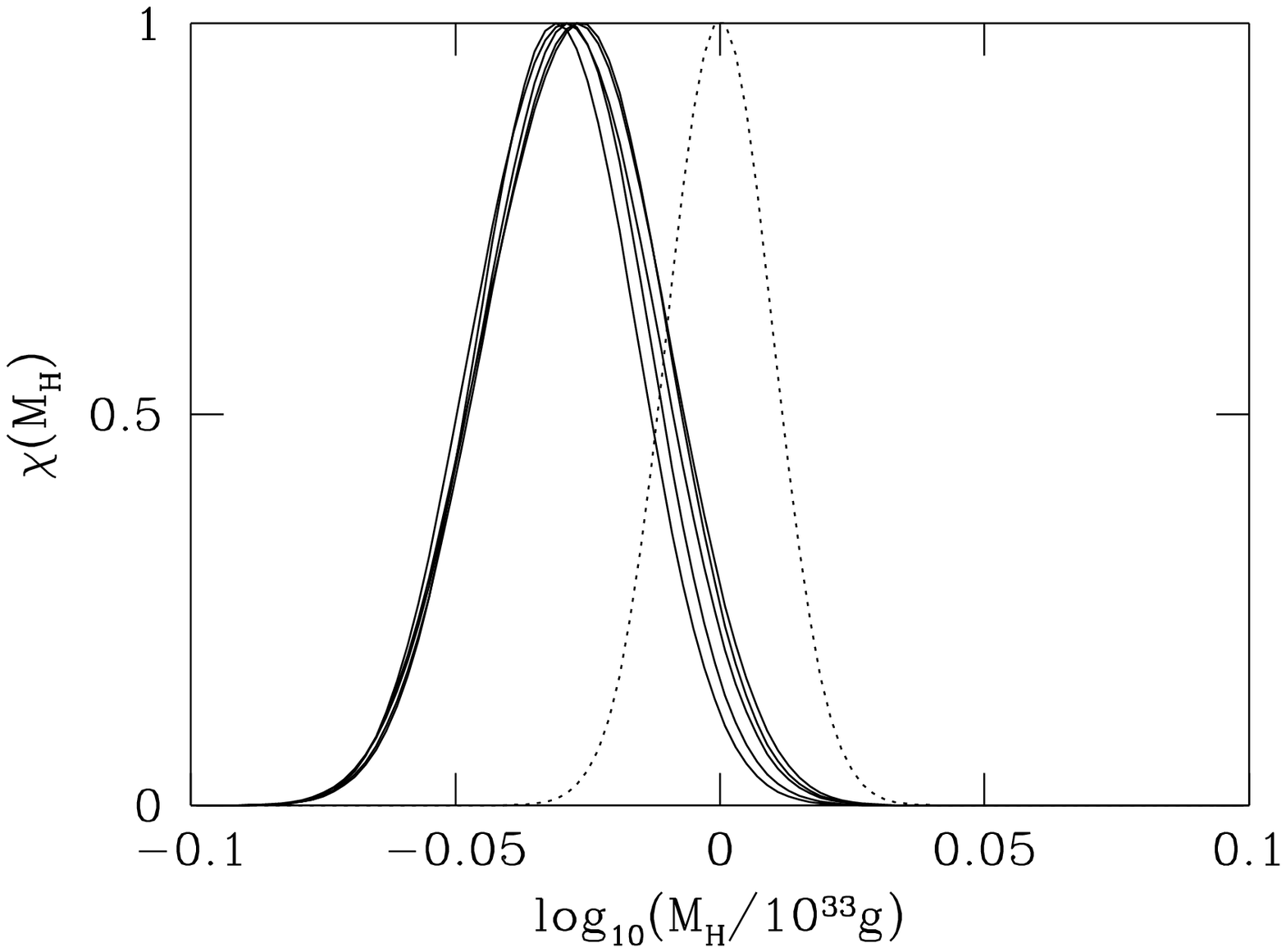}\\
\caption[sfixmbh]{\label{sfixmbh} The smoothed distributions of the
actual masses, $\chi(M_{{\rm BH}})$, (upper panel) and horizon masses,
$\chi(M_{{\rm H}})$ (lower panel) of PBHs formed due to a spike in the
density perturbation spectrum with $\Sigma= 1\times 10^{9} {\rm
Mpc}^{-1}$ and (from right to left on the right-hand side of each
diagram) $C=1.75 \times 10^{10}, 1.5 \times 10^{10}, 1.25 \times
10^{10}, 1 \times 10^{10}$ and $7.5 \times 10^{9} {\rm Mpc}^{-1}$. The
dotted line in the upper panel shows the Niemeyer and Jedamzik mass
function evaluated for $\sigma(10^{33}{\rm g})=0.053$, and in the
lower panel shows a smoothed delta function centered on $M_{{\rm
H}}=10^{33}$g.}
\end{figure}

Fig.~\ref{sfixmbh} shows the same distributions for $\Sigma= 1 \times 10^{9}
{\rm Mpc}^{-1}$ and a range of $C$ values.  As $C$ is decreased the centre of
the PBH mass function tends to that of the NJMF.  The distribution of the
horizon masses at which the PBHs form is independent of $C$ and centered at
$M_{{\rm H}}< 10^{33}$g.

The horizon masses at which the PBHs form are typically smaller than
$10^{33}$g, since whilst ${\cal P}_{{\cal R}}$ has a spike centered at
this scale, $\sigma(M_{{\rm H}})$ carries on increasing, as $M_{{\rm
H}}$ is decreased, until ${\cal P}_{{\cal R}} \ll C$ on that scale.
By increasing the scale on which the spike is centered we could tune
the horizon mass distribution to be precisely centered at $M_{{\rm
H}}=10^{33}$g.

The PBH masses are typically larger than given by the NJMF with 
$\sigma(10^{33}){\rm g}=0.053$, despite
being formed at smaller $M_{{\rm H}}$.  This is because our spiky
density perturbation spectra all have $\sigma(M_{{\rm H}})\sim 0.2$,
in order to produce a larger number of PBHs than is allowed by
observations (as is discussed above), leading to larger values of
$\delta(M)$ and hence $M_{{\rm BH}}$.

\begin{figure}
\centering
\leavevmode\epsfysize=6.3cm \epsfbox{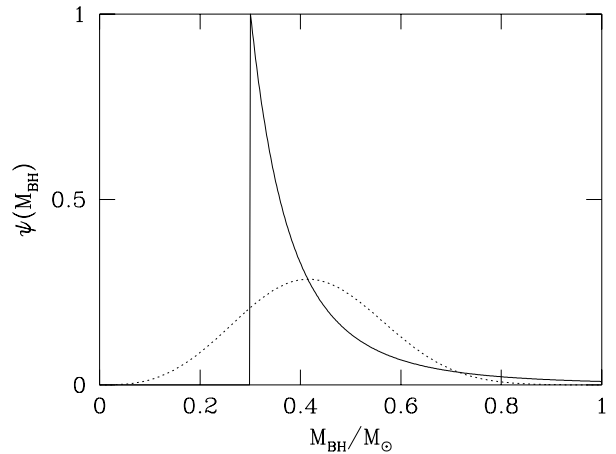}\\
\caption[macnj]{\label{macnj} The observed distribution of the MACHOs,
given by Eq.~(\ref{machdist1}), is shown as the solid line and the
Niemeyer and Jedamzik mass function, given by Eq.~(\ref{njmf}), as the
dashed line. We fix $M_{{\rm H}}=2.05 \times 10^{{33}}$g so that the
mass functions have the same mean. The curves are chosen to have the
same integrated mass fraction and the same mean.}
\end{figure}

Since the spread of $M_{{\rm H}}$ at which the PBHs form will be
extremely small if the amplitude of the spike, and hence $\sigma(M)$,
is reduced so that the number of PBHs produced satisfies the
constraint $\Omega_{({\rm PBH,0})}<1$, then the PBH mass function will
be the same, to a good approximation, as that found by assuming that
all the PBHs form at a single horizon mass.  We therefore compare, in
Fig.~\ref{macnj}, the observed MACHO mass function, $\psi(M)$, with
that determined by Niemeyer and Jedamzik, Eq.~(\ref{njmf}).  Defining
$\tilde{\psi}(M) {\rm d}(\ln{M})$ to be the mass fraction in a
logarithmic mass interval ${\rm d}(\ln{M)}$, then $\tilde{\psi}(M)=
\psi(M) M$. Since ${\rm d}N/{\rm d}(\ln{M}) \propto \tilde{\psi}/M$,
then $\psi(M_{{\rm BH}}) \propto {\rm d}N/{\rm d}(\ln{M_{{\rm BH}}})$.
That is to say, the mass fraction per unit mass interval is just
proportional to the number density per logarithmic mass interval.  We
adjust the horizon mass at which the PBHs are assumed to form at to
$M_{{\rm H}}=2.05\times 10^{33}$g, so that the PBH mass distribution
has the same mean mass as the MACHO distribution. The mass
distributions are normalized to give the same total mass fraction when
integrated.

The PBH mass distribution is much broader than that fitted to the
observed microlensing events; its full width at half maximum is six
times larger (assuming the mass functions are normalized to the same
mean). Because the derivation assumes all black holes form at the same
horizon mass, it is the narrowest possible mass function which can
arise if the critical collapse hypothesis is correct.  The mass
function is almost symmetrical, extending to $M_{{\rm BH}} \ll
M_{{\rm H}}$.  Currently there is a robust limit, due to the absence
of short-duration microlensing events, that less than 20 percent of
the dark matter in the halo of the galaxy can be in low mass MACHOs
($10^{-4}M_{\odot}< M< 0.03 M_{\odot}$) \cite{ma:cho}, but the PBH
mass function is easily consistent with that. Whether or not the width
of the predicted mass function is in agreement with the observed
microlensing events, which so far have been fitted with much narrower
functions, is a much more pertinent question. As the duration of
microlensing searches increases, and further microlensing events are
observed, the MACHO mass function will be determined much more
accurately, and if the MACHOs are PBHs then the true MACHO mass
function will be considerable wider than the sharply-peaked mass
functions which have been fitted to the current data.

\section{Conclusions}

Critical gravitational collapse, and in particular the scaling
relation for the black hole mass, has an astrophysical application to
PBH formation.  We have used the excursion set formalism to determine
the PBH mass function, when formation on a range of mass scales is
taken account of, for two different types of power spectra.  The first
is power-law density perturbation spectra, and the second is flat
spectra with a spike on a given scale.  In both cases we find that as
the parameters are adjusted so that the abundance of PBHs decreases,
the mass function tends towards that found
by Niemeyer and Jedamzik under the assumption that all the PBHs form
at a single horizon mass.  There are tight observational limits on the
abundance of PBHs, and for power spectra which satisfy these
constraints the assumption that all PBHs form at the same horizon mass
is a good approximation.

As an application, we then compared the Niemeyer and Jedamzik PBH mass
function with that of the MACHOs found from microlensing observations.
The PBH mass distribution is considerably broader than the
sharply-peaked simple mass distributions which have so far been fitted
to the observed microlensing events.  If the MACHOs are indeed PBHs,
then future microlensing searches should unveil the broad spread of
the PBH mass distribution.

\section*{Acknowledgments}

A.M.G.~was supported by PPARC and A.R.L.~in part by the Royal
Society. We thank Bernard Carr, Karsten Jedamzik and Jens Niemeyer for
useful discussions.  A.M.G.~acknowledges use of the Starlink computer
system at QMW.

\end{document}